\documentclass[preprint]{elsarticle}
\usepackage{graphicx}
\usepackage{geometry}
\usepackage{gensymb}
\usepackage{amsmath}
\usepackage{physics}
\usepackage{lipsum}
\usepackage{xcolor}
\usepackage{url}
\usepackage{lineno}
\usepackage[T1]{fontenc}
\usepackage[utf8]{inputenc}
\usepackage[skip=2mm]{parskip}
\usepackage{mathptmx}
\usepackage{hyperref}
\usepackage{natbib}
\usepackage{lineno}

\newcommand{\Xmax}{$\mathrm{X_{max}}$}
\newcommand{\FMmax}{$\mathrm{FM_{max}}$}
\newcommand{\FMwidth}{$\mathrm{FM_{FWHM}}$}
\newcommand{\Cerenkov}{\v{C}erenkov }

\begin{document}
\title{Air shower development through the time dependence of its induced electric field}

\author[1,2]{Beatriz de Errico\corref{cor1}}
\author[2,3]{Charles Timmermans}

\cortext[cor1]{Email: \href{mailto:beatrizspe@pos.if.ufrj.br}{beatrizspe@pos.if.ufrj.br}}

\affiliation[1]{organization={Instituto de Física, Universidade Federal do Rio de Janeiro},
            addressline={Cidade Universitária - Ilha do Fundão}, 
            city={Rio de Janeiro},
            state={RJ},
            country={Brazil}}

\affiliation[2]{organization={Institute for Mathematics, Astrophysics and Particle Physics, Radboud Universiteit},
            city={Nijmegen},
            country={the Netherlands}}

\affiliation[3]{organization={Nikhef, National Institute for Subatomic Physics},
            city={Amsterdam},
            country={the Netherlands}}

\begin{abstract}
Ultra-high energy cosmic rays impinge on the atmosphere and induce air shower cascades, in which huge numbers of particles are produced. By travelling through the Earth's atmosphere and magnetic field, these particles create a noticeable effect on the electric field at the surface. In this article, we demonstrate that parameters of the shower longitudinal development and primary mass can be inferred from mapping the time dependence of the observed electric field to the emitted electric field as a function of slant depth along the shower axis.
\end{abstract}


\maketitle

\section{Introduction}

Since the first measurements in 1965, cosmic ray induced air showers have been known to create disturbances in the Earth's electric field that can be detected using antennas \cite{Jelley:1965nue}. These disturbances are mainly the result of the interaction of the air shower particle cascade with the Earth's magnetic field throughout the air shower development in the atmosphere, referred to as the geomagnetic emission mechanism in the literature \cite{KahnLerche}. In addition, excess of negative charges at the shower front produce Askaryan radiation \cite{Askaryan:1961pfb}, which in air is a secondary effect. Both mechanisms create coherent radiation in the radio frequency range, from a few to 500~MHz, with different polarization signatures. 

Measurements of air shower-induced radiation have been interpreted and used in reconstruction methods \cite{frankSH,Jhansi:2024trc,Schluter:2022mhq,Huege:2016veh}, allowing us to derive air shower properties, such as its geometry, the energy contained in it, and the atmospheric depth at which particle production reaches its maximum, \Xmax. With these observables, the characteristics of the primary particle that generated the air shower are inferred \cite{Gaisser_Engel_Resconi_2016}. Beyond that, the calorimetric character of radio-wave detection opens the possibility of probing the longitudinal development of air showers through the atmosphere, which so far has only been done using fluorescence telescopes \cite{PierreAuger:2018gfc}.

A longitudinal reconstruction of an air shower through its radiation measurements requires knowledge of the time it takes for electromagnetic signals to travel from their emission points to different observers. The interferometric reconstruction method \cite{Schoorlemmer:2020low,Schluter:2021mjt} uses differences in travel times between observers to obtain a signature per atmospheric depth that can be correlated with the maximum of the longitudinal profile, but not with its shape. The approach described in \cite{Jhansi:2024trc} uses the times of maximum amplitudes of the detected electric fields to fit a shower front, which is employed to estimate an effective point source on the shower axis of the fluence at each observer. It is demonstrated that by using multiple measurements of a single event, the obtained fluence as a function of depth profile is correlated to the shower longitudinal profile and can be used to estimate \Xmax. 

In this article, we also propose an approach to reconstruct a longitudinal radio-wave emission profile, however we rely solely on the absolute radiation travel times. For a given event, we only use the air shower geometry, the observer position, and its measured electric field time series, all easily obtained at any air shower-experiment, e.g. \cite{AugerPrime,Sagawa:2013pya,whitePaperGRAND}, and profit from the fact that all observers see the emission from all points along the cascade. Using at least a single antenna, we map each time bin of its observed electric field time series to an atmospheric depth slice on the shower axis. With it, we create a field mapping profile that can also be correlated with the longitudinal particle profile. Beyond estimating \Xmax ~and primary particle type, we expect the approach to possibly be a new method to probe the development of the air shower cascade and its radio-wave emission mechanisms, as well as perform cosmic ray mass composition analysis.

\begin{figure*}[t]
	\centering
	\includegraphics[width=.7\textwidth,trim={4.6cm .2cm 6.5cm 0cm}, clip]{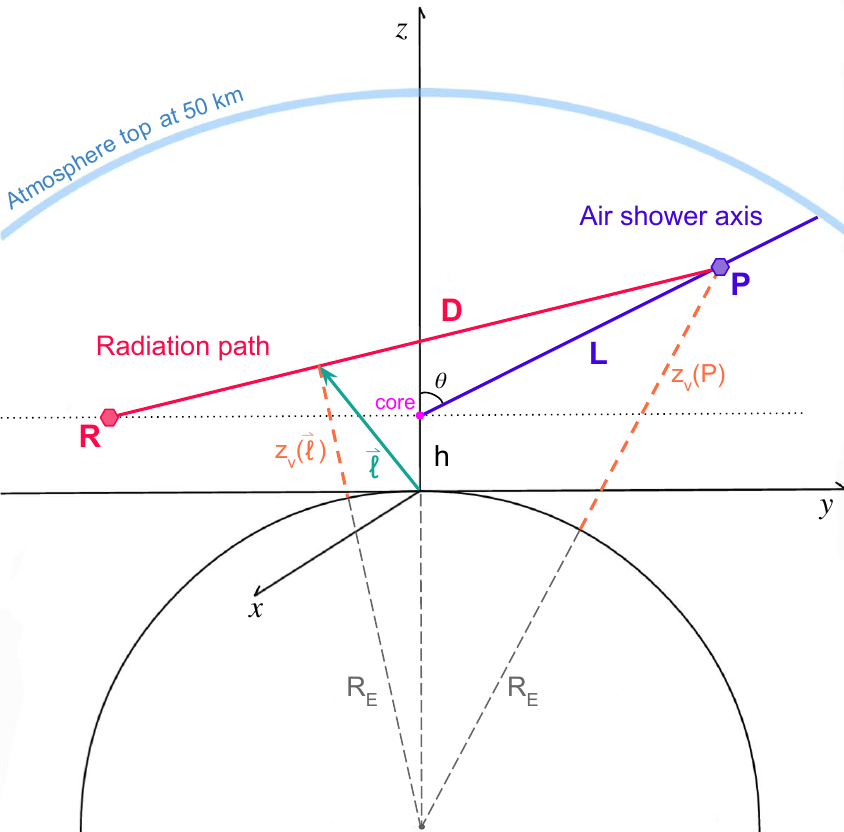}
	\caption{Simplified geometry of an air shower with its complete development contained along the shower axis. The top of the atmosphere is set at 50\,km. The shower axis is at an angle $\theta$ w.r.t. the zenith at the core position. The detection plane is at an altitude h above sea-level. Radiation emitted from point P, located on the shower axis at distance L from the core, is observed at point R, at a distance D to the emission point. The radiation path is parametrised by $\vec l$.}
	\label{fig:Geometry}
\end{figure*}

\section{Method}

This method relies on the assumptions that radiation is primarily emitted on or very near the air shower axis and that it travels in straight paths from the point of emission to the observer, reasonable approximations as discussed in Ref. \cite{VandenBroeck:2024dec, WERNER2008393}. Using the known air shower geometry, zenith and azimuth angles and core position\footnote{Where the particle cascade hits the surface of the Earth on the shower axis.}, we determine the shower axis from a starting altitude of 50\,km above the surface of the Earth. In Figure \ref{fig:Geometry} we illustrate the geometry considered for the calculations. For a given observer position R, we calculate the time that the radiation emitted at different points P along the axis takes to reach R, relative to the time the shower particles take to travel from each position P to the core. 

The calculation of the absolute radiation travel time for one position P along the axis, $T_{PR_{abs}} $, is given by:
\begin{equation}
    T_{PR_{abs}} = T_{PR}-T_{Pcore} = \int_{D}{n(z_v(\vec l))\over c} d \vec l - {L\over \left<\beta\right> c}
    \label{eq:dt}
\end{equation}
where $c$ is the speed of light in vacuum. The particles travel time, $T_{Pcore}$, depends on the average velocity of the fastest particles near the shower axis $\left<\beta\right>c$, in which the approximation of $\left<\beta\right> = 1$ is reasonable, in accordance with Ref. \cite{Alvarez-Muniz:2010hbb}. The radiation travel time, $T_{PR}$, depends on the refractive index $n$ along the path $D$. Furthermore, $n$ is a function of the height above the Earth's surface, which we call the vertical height, thus $n = n(z_v)$. 

In the limit of air showers with zenith angles higher than $60\degree$, the discrepancy between height and vertical height due to curvature effects becomes significant and must be taken into account in the refractive index estimation.
In Figure \ref{fig:Geometry}, we illustrate the vertical height, $z_v$, of a point along the radiation path characterised by the cartesian coordinates $\vec l = (l_x,l_y,l_z)$. We can calculate $z_v$ by the following relation:
\begin{equation}
    z_v(\vec l) = \sqrt{(l_z + R_E)^2 + l_x^2+l_y^2} - R_E,
    \label{eq:zv}
\end{equation} 
where $R_E$ is the radius of the Earth and the vertical coordinate $l_z$ includes the height $h$ of the core position above sea-level. 
The refractive index is parametrized by an exponential function:
\begin{equation}
	n(z_v) = 1 + A e^{-B z_v},
	\label{eq:n}
\end{equation}
with $A = 325 \cdot 10^{-6}$ and $B=0.1218 \cdot 10^{-3}$\,m$^{-1}$, as used in the air shower simulation package ZHaireS \cite{Alvarez-Muniz:2010hbb}. When combining equations \ref{eq:zv}, \ref{eq:n}, and \ref{eq:dt}, the obtained path integral is not analytically solvable. Therefore, we use the 3D-coordinates of the considered observer R and emission position P to parametrise the radiation path and perform a numerical integration of $T_{PR}$. With this approach, we maintain computing efficiency without requiring an estimation of an effective refractive index along the radiation path.

Using this framework, we can calculate an atmospheric depth time series for various observers. We slice the shower axis into fixed segments of 10\,m and, for each of them, we use equation \ref{eq:dt} to calculate the absolute radiation travel time to all observers on the surface. This provides the time-basis for the electric field time series seen by each observer.  In the following, we assign each time bin of the electric field time series to a range in atmospheric slant depths, using the standard US atmospheric model \cite{USstandard} to match the ZHaireS air shower simulations. With this method, we obtain a cumulative electric field amplitude mapping profile onto the depth traversed by the air shower, which we refer to as the field mapping profile. 

The precision of this method is determined by the relationship between the time intervals at which the electric field is measured and the amount of air that the cascade has traversed between the corresponding emission points. In general, the larger the distances between the observers and the shower axis, the more the radiation signals spread over time, resulting in a more detailed image of the development of the air shower. The interplay between the propagation speeds of the cascade particles and their emitted radiation leads to interesting relativistic effects. At the Cherenkov angle w.r.t. the shower axis in air, a ring-like region is delimited where the entire cascade emission arrives in a short time interval, enhancing coherence \cite{PhysRevD.86.123007}. This is referred to as the \Cerenkov region in literature, and observers close to or within this region will have a coarser picture of the shower development.

In this paper, we demonstrate the validity of the outlined method using air shower simulations. We obtain field mapping profiles from the radio-wave emission as seen by observers outside the \Cerenkov region and correlate their characteristics with the shape of the air shower longitudinal profile. We also compare the field mapping profile shape obtained from different primary particle types and asses the potential of performing mass composition studies with the described method.

\begin{figure}[t]
    \centering
    \includegraphics[width=\linewidth,trim={0cm 0cm 0cm 0cm}, clip]{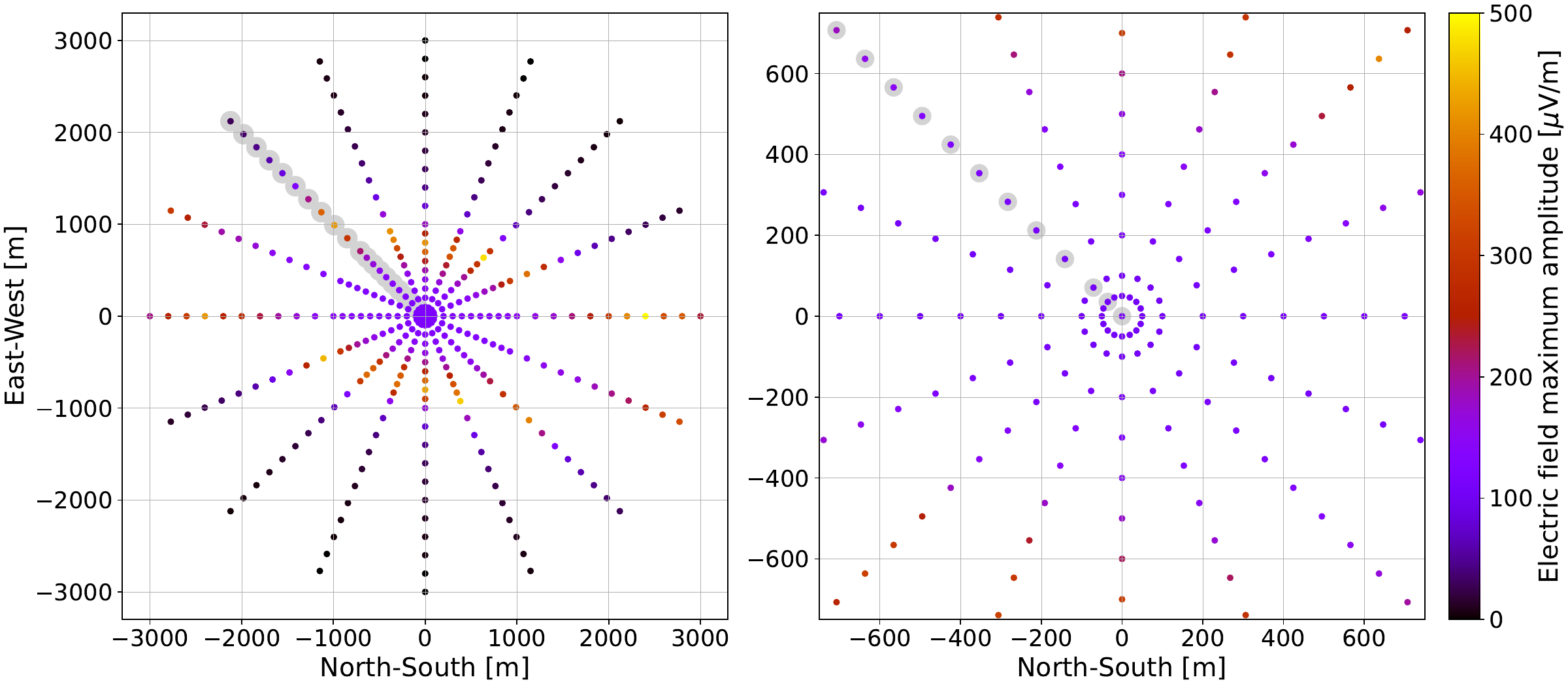}
    \caption{\textbf{Left:} Detector array on the ground plane used for the simulations. The layout consists of a star-shape with 16 spokes and 21 concentric rings. The colour scale shows the maximum electric field intensity for each observer, representative of the radio-wave ground footprint of a 1\,EeV proton-induced shower, with a zenith angle of $\theta = 80\degree$, and an azimuth angle of $\phi=345\degree$ w.r.t. the magnetic North. The observers on the spoke at $120\degree$ from north, highlighted in grey, are used to exemplify the method in the following sections. \textbf{Right:} A detailed view of the denser setup region near the shower core.}
    \label{fig:simusetup}
\end{figure}

\section{Implementation on air shower simulations}

We used the AIRES package version 19.04.10 \cite{Sciutto:1999jh} to simulate air showers. The ZHaireS extension, version 1.0.30c \cite{Alvarez-Muniz:2010hbb}, is adopted to calculate the emission and propagation of electromagnetic waves from the particle cascade. We chose the pre-defined location of Malargüe, Argentina, corresponding to the site of the Pierre Auger Observatory \cite{PierreAuger:2015eyc} at -35.2 degrees latitude, -69.20 degrees longitude, and altitude of 1425\,m above sea-level. We also used the Geomagnetic field at Malargüe, with a magnitude of 23.48\,$\mu$T. Its inclination is $-37.29\degree$ and its declination $0.23\degree$. Note that all coordinates in our simulations are with respect to the magnetic North.  

For this location, we simulated 240 proton-initiated showers at an energy of 1\,EeV, a zenith angle of $80\degree$, and at azimuth angles ranging from 0 to $360\degree$ spaced by 15 degrees. For each combination of zenith and azimuth angles, we generated 10 events. SIBYLL2.3d \cite{Riehn:2019jet} is used to describe the hadronic interactions, the standard US atmospheric model \cite{USstandard} is used to model the atmosphere, and the thinning is set to relative $10^{-6}$. As we present a first proof of principle study, we focus on near horizontal showers where the effects of the method are more prominent. We also simulated 240 iron-initiated showers at the same energy, zenith, and azimuth angles range. For further inspection on primary particle dependence, we simulated an extra set of 240 proton-initiated showers, where we set the first interaction depth to be the same as the corresponding iron shower with the same geometry. Applying the method in a realistic scenario would also include the detector response and noise interference. This has to be carefully considered for each  air shower experiment.

\begin{figure}[t]
	\centering
	\includegraphics[width=.9\textwidth]{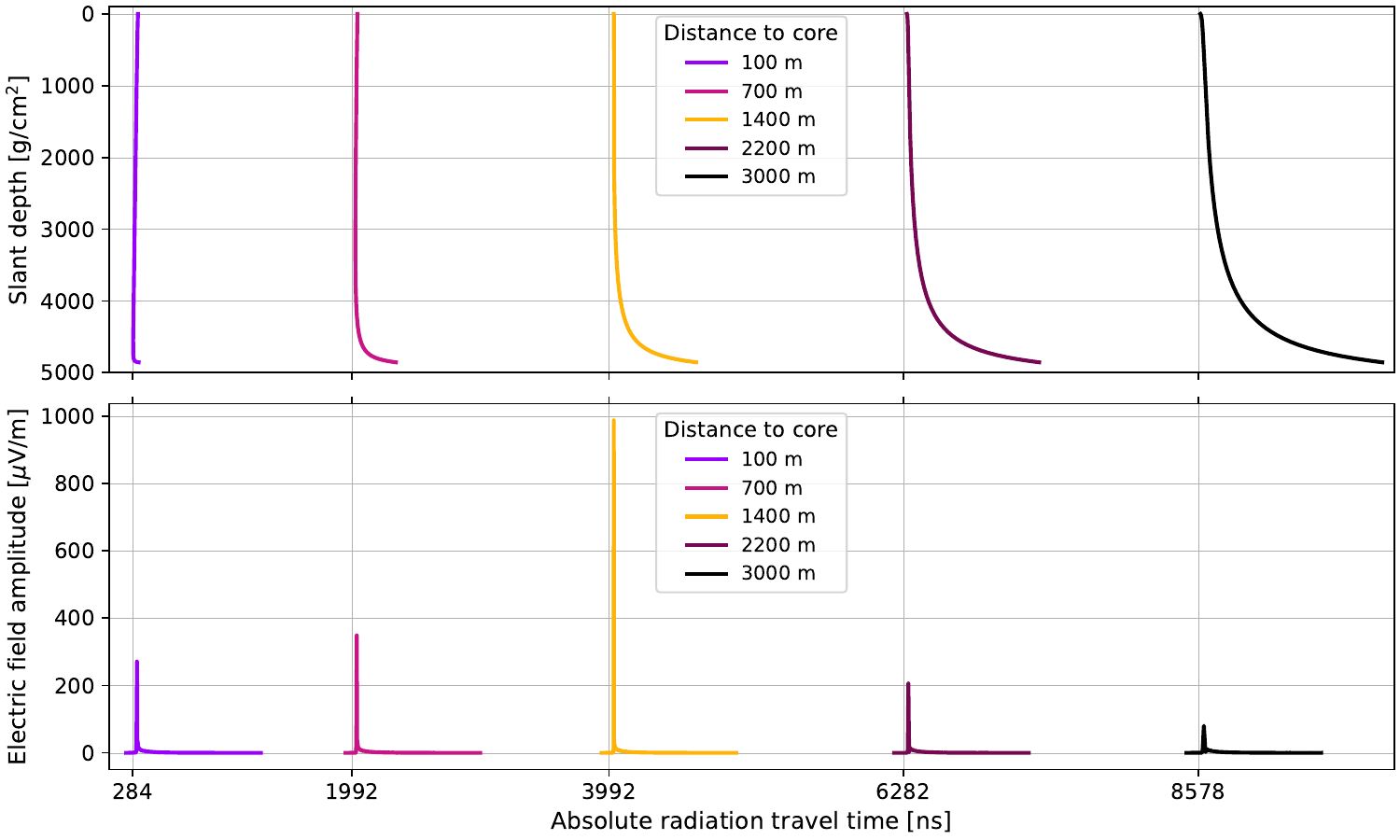}
	\caption{\textbf{Top:} $T_{R_{abs}}$ as a function of the slant depth, calculated for five observers located along the highlighted spoke in Figure \ref{fig:simusetup}, and an air shower geometry of $\theta = 80\degree$ and $\phi=345\degree$. The observers were chosen for their positions w.r.t. the \Cerenkov region: two outside and close to the core (at 100 and 700\,m), one within the \Cerenkov region (at 1400\,m), and two outside and far away from the core (at 2200 and 3000\,m). \textbf{Bottom:} The simulated electric field time series of a proton-induced event for the five selected observers. The marked labels in the horizontal axis correspond to the time the electric field becomes non-zero at each observer, representative of when the first signal arrives at each location.}
	\label{fig:Timing}
\end{figure}

In Figure \ref{fig:simusetup}, we show the layout of the detection setup. The observers were positioned at the origin and in 21 concentric rings around it, each with 16 detectors. The spacing of the rings is chosen to be 50\,m up to a radius of 100\,m. Afterwards the spacing is increased to 100\,m up to a radius of 1000\,m and finally they are separated by 200\,m up to a radius of 3000\,m. For each air shower event, we set the core position at the origin and simulated the measured electric field at each observer, using a time window of 1.5\,$\mu$s and resolution of 1\,ns. The setup choice is optimized to investigate the dependence of the induced electromagnetic signals with both the angle w.r.t. the shower axis and the distance to the core. Note that we defined this star-shaped layout in the ground reference frame, as opposed to the usual shower-plane reference frame adopted by the community. This choice was made because the method relies on the ground plane geometry between the detector position and the emission region, which is reflected in the electric field time series amplitude distribution. This becomes especially relevant for inclined showers.

Without further assumptions, we applied the method to the simulation set produced. Using equation \ref{eq:dt}, we calculated the absolute radiation travel time as a function of slant depth for each observer, $T_{R_{abs}}(X)$. In the upper panel of Figure \ref{fig:Timing}, we show $T_{R_{abs}}(X)$ for five observers located along the highlighted spoke in Figure \ref{fig:simusetup}, which have strategic positions w.r.t. the \Cerenkov region. In the lower panel of Figure \ref{fig:Timing}, we show the simulated electric field time series at each of these observers, in which t$_0=0$\,ns corresponds to the time the particle cascade reaches the core position. Observers further away from the emission points detect the electromagnetic waves later in time. This effect is commonly used to reconstruct the direction of propagation of the air shower, and thus also of the incoming cosmic ray \cite{PierreAuger:2018pmw}.

\begin{figure}[t]
    \centering
    \includegraphics[width=\linewidth]{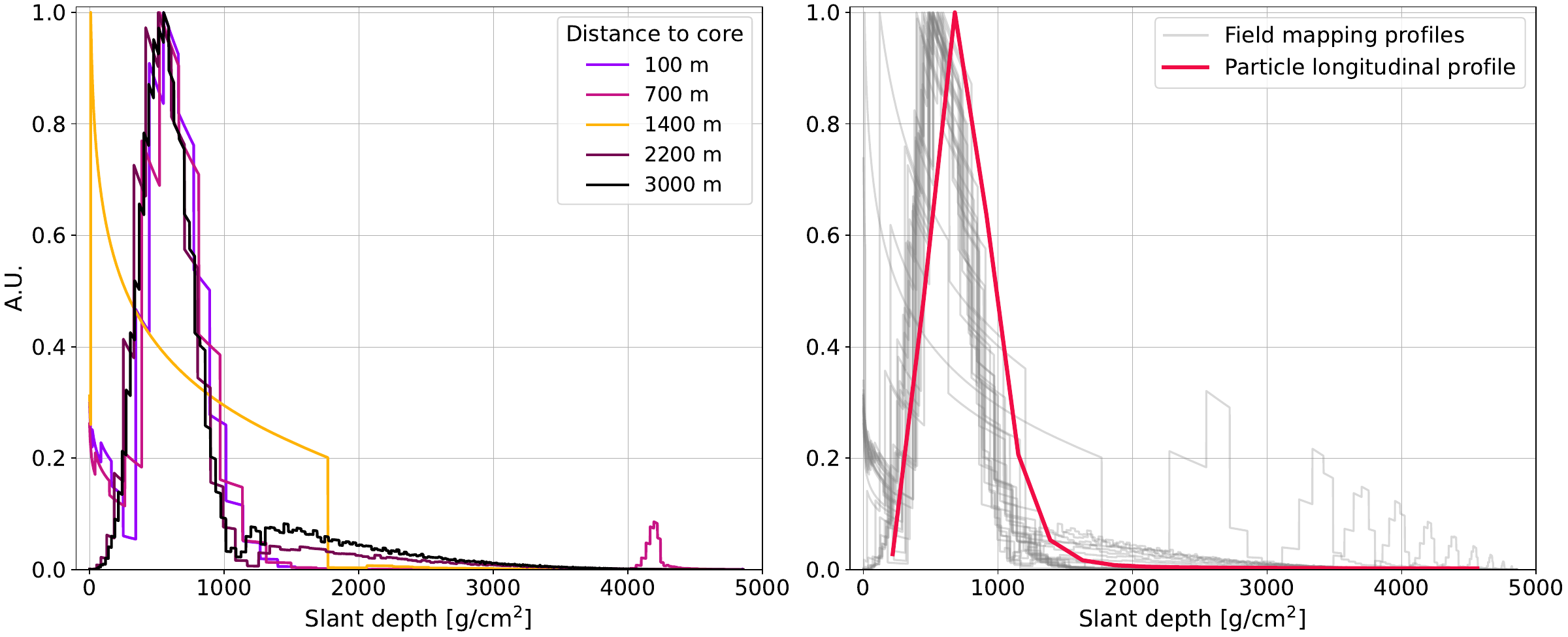}
    \caption{\textbf{Left:} Mapping profiles for detectors positioned outside the \Cerenkov region and close to the core (100 and 700\,m), within the \Cerenkov region (1400\,m), and outside the \Cerenkov and far away from the core (2200 and 3000\,m). All profiles are normalized from their maximum amplitude, so to better illustrate their similarities in shape. \textbf{Right:} The normalized mapping profile shapes correlate with the normalized longitudinal particle profile of the air shower. The sharp particle profile is due to conversion from vertical to slant depth.}
    \label{fig:mapdepth}
\end{figure}

A comparison between the upper and lower parts of Figure \ref{fig:Timing} demonstrates that the two assumptions made for radiation emission and propagation seem to hold. The time range for the slant depth mapping and the electric field time series agree and also indicate that the bulk of the radio-wave emission originates from the region shallower than 1000\,g/cm$^2$ for all observers. In addition, the timing per slant depth curves depict the relativistic effects of the particle cascade traveling faster than the emitted radiation in air. For the two antennas closest to the core, we see that the radiation emitted deeper in the atmosphere arrives at the observer before the radiation emitted higher up. This effect is reversed on the two observers further away. The transition is seen at the observer within the \Cerenkov region, in which most of the emissions arrive within one or two time bins.

For each observer R, we combine the top and bottom parts of Figure \ref{fig:Timing} to map the electric field time series $E_R(t)$ onto the slanted depth timing $T_{R_{abs}}(X)$, i.e.:
\begin{equation}
    E_R(t=T_{R_{abs}}(X)) = E_R(X),
\end{equation}
profiting from the imposed t$_0=0$\,ns at the core position for both time references. In the left of Figure \ref{fig:mapdepth}, we show the resulting field mapping profiles for the same five selected observers, a cumulative distribution of all the electric field amplitudes that originate from the same slant depth bin. We re-scale each slant depth bin contribution with its distance to the observer R, to account for the decrease in the electric amplitude due to propagation. After re-scaling, the profiles are normalized such that their maximum amplitudes are set to one.

The resulting mapping is not linear, since more than one electric field time bin can originate from the same depth, as seen in the timing curves in Figure \ref{fig:Timing}. For observers outside the \Cerenkov region and close to (far away from) the core, there is a unique mapping from the time of arrival to atmospheric depth, where the later (earlier) part of the shower arrives first. It is interesting to note that this does not significantly affect the shape of the main peak. Within the \Cerenkov region, as seen in the yellow curve in the left panel of Figure \ref{fig:mapdepth}, there is a pile up of amplitudes in the first depth bins, due to coincident arrival times. The resulting mapping gives a rougher view of the shower profile, and we therefore discard observers within the \Cerenkov region from the analysis.

In the right panel of Figure \ref{fig:mapdepth}, we show the normalized field mapping for all observers along the highlighted spoke in Figure \ref{fig:simusetup} superimposed on the longitudinal profile of electrons and positrons in the simulated air shower particle cascade. We also normalize the particle profile by its maximum amplitude, in order to facilitate shape comparisons with the mapping profiles. For observers outside the \Cerenkov region, the overall shape of the field mapping profile seems to be nearly independent of the observer position. The maximum of the field mapping profile, \FMmax, appears consistently before the cascade reaches \Xmax, with some small fluctuations. This corresponds to the region of maximum charge variation in the cascade, which is in agreement with the emission mechanisms described in the literature \cite{Scholten:2017tcr}. Overall, we can confirm that, outside the \Cerenkov region, all observations of the local electromagnetic field can be used to infer the longitudinal profile of the radiation.

\begin{figure}[t]
    \centering
    \includegraphics[width=.6\linewidth,trim={1.3cm 0.5cm 1.5cm 1.5cm}, clip]{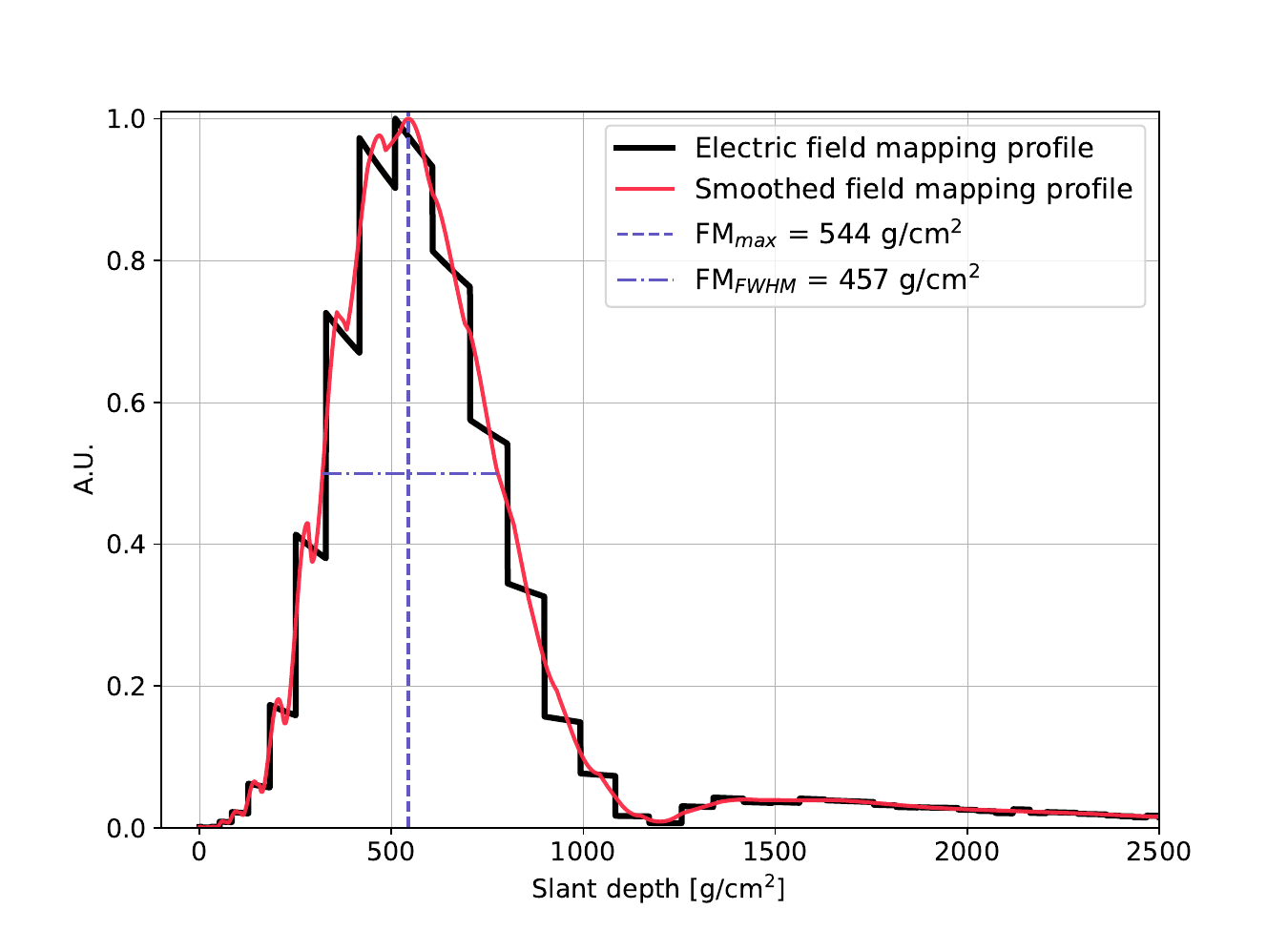}
    \caption{Original harsher field mapping profile, in black, and its smoothed output, in red. The profile shape parameters, \FMmax ~and \FMwidth, are extracted from the smoothed profile.}
    \label{fig:smooth}
\end{figure}

Next, we systematically characterise the shape of the field mapping profiles. To avoid harsh shapes due to the time binning of the electric field time series, we apply a Savitzky–Golay filter \cite{savgol_filter} to smooth the field mapping profile. This filter was chosen due to its good performance in maintaining the original shape of the input. In Figure \ref{fig:smooth}, we show the original and smoothed field mappings for the observer at $2200$\,m from the core. From the smoothed mappings, we can more reliably extract the slant depth at which the profile reaches maximum amplitude, \FMmax, as well as its full width half maximum, \FMwidth. We approximate the particle longitudinal distribution by a Gaisser-Hillas (GH) distribution \cite{Andringa:2011zz}:
\begin{equation}
    N_{GH}(X)=\left(1 + R\frac{(X-X_{max})}{L}\right)^{R^{-2}}\text{exp}\left(-\frac{(X-X_{max})}{LR}\right),
\end{equation}
fitting the \Xmax, L, and R parameters. To search for correlations between the profile shapes, we compare the obtained \Xmax ~and $L$ of the GH with the \FMmax ~and the \FMwidth ~of the smoothed field mappings.

\section{Results}

\begin{figure}[t]
	\centering
	\includegraphics[width=\linewidth]{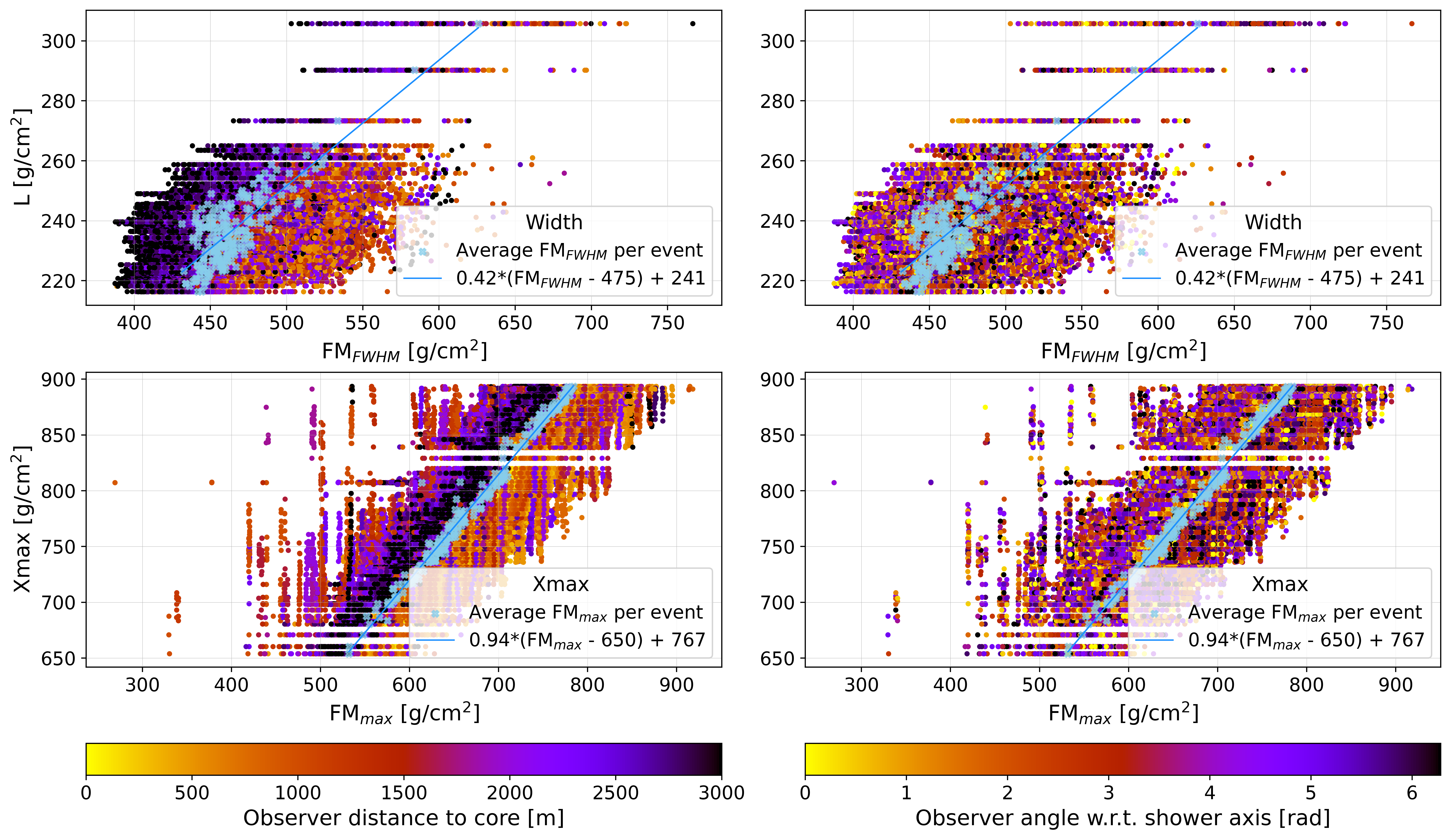}
	\caption{\textbf{Top:} the correlation between L, the width parameter of the particle longitudinal profile, and \FMwidth ~for proton-induced air showers. \textbf{Bottom}: The correlation between Xmax and \FMmax ~for proton-induced air showers. \textbf{Left:} The color code follows the distance of the observer to the shower core, in the ground reference frame. \textbf{Right:} The color code shows the angle w.r.t. the projection of the shower axis in the XY plane of the observer spoke, also in the ground reference frame. In all panels, the light blue markers show the average field mapping parameters for each event.}
	\label{fig:FMworld}
\end{figure}

\begin{figure}[h!]
	\centering
	\includegraphics[width=\linewidth]{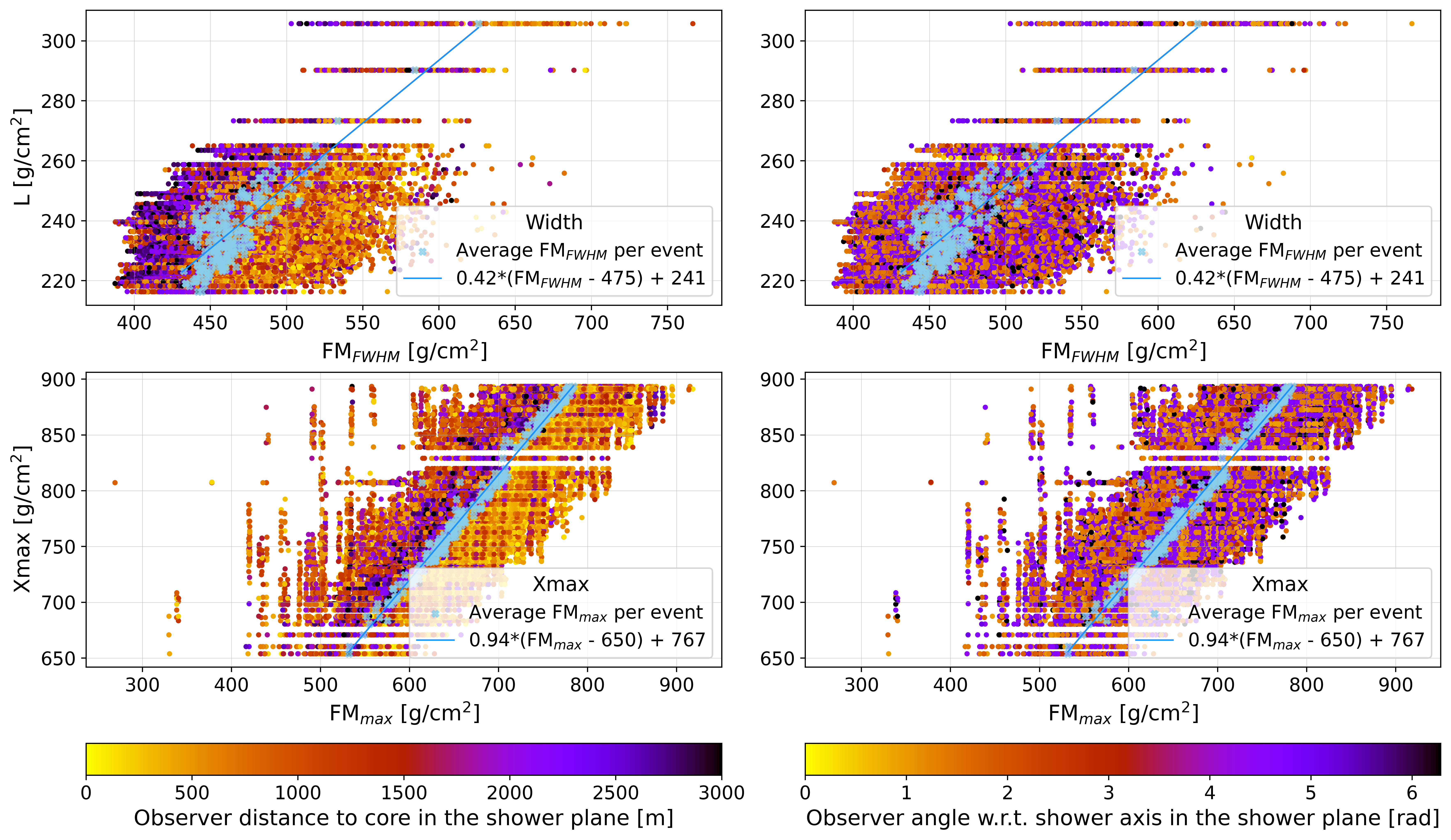}
	\caption{The same correlation distribution between L and \FMwidth, on the first row, and \Xmax ~and \FMmax, on the second row, as for Figure \ref{fig:FMworld}. However, the color code shows the observer distance to the core and the angle w.r.t. shower axis in the shower plane referential.}
	\label{fig:FMSP}
\end{figure}

For each of the 240 proton-initiated events, we obtained the field mapping profile for all observers outside the \Cerenkov region and their corresponding shape parameters \FMmax ~and \FMwidth, as well as the \Xmax ~and L for the GH fit of the particle longitudinal profile. In the first row of Figure \ref{fig:FMworld}, we show the correlation between L and \FMwidth; color coded by the distance of the observer to the shower core in the first column and by the angle with respect to the projection of the shower axis in the XY plane in the second column. In the second row of Figure \ref{fig:FMworld}, we show the correlation between \Xmax ~and \FMmax, using the same color coding.

To better assess the correlation between the profiles, we calculated the average \FMwidth ~and \FMmax ~over all observers of each simulated event, shown by the blue markers in Figure \ref{fig:FMworld}. The relation between the average width of the field mapping profiles and the width of the particle distribution is given by: 
\begin{equation}
    L = 0.42~(\text{FM}_{\text{FWHM}}-475\,\text{g/cm}^2) +241\,\text{g/cm}^2.
    \label{eq:width_fit}
\end{equation}
We fit the first order polynomial in the average \FMwidth ~subtracted by the median distribution value of 475\,g/cm$^2$. In doing so, we ensure that the uncertainty of 0.5 g/cm$^2$ on the fit of the linear coefficient corresponds to a good estimate of the uncertainty on the obtained value for L in the whole region of interest. In addition, the relation between the averaged \FMmax ~and \Xmax ~is given by: 
\begin{equation}
    \text{X}_{\text{max}} = 0.94~(\text{FM}_{\text{max}}-650\,\text{g/cm}^2) +767\,\text{g/cm}^2.
    \label{eq:Xmax_fit}
\end{equation}
The median value of \FMmax ~is close to 650\,g/cm$^2$, and using the same procedure as for \FMwidth, we obtain an uncertainty of 0.6 g/cm$^2$ for our \Xmax~estimate. It should be stressed that this result only holds when using the average values of \FMwidth ~and \FMmax ~taken over all simulated observers. Furthermore, full knowledge on the electric field and no noise contribution are assumed.

From the color scale on the left side of Figure \ref{fig:FMworld}, we see a clear radial dependence correlation of both the widths and the maxima. However, from the right side of this Figure, no obvious correlation with the angle of the spoke in ground coordinates can be observed. In Figure \ref{fig:FMSP}, we show the same correlations; however, the color code represents the distances and angles in the shower plane. The conclusions are very similar, as also in the shower plane, the distance to the core affects the correlation between the widths and maxima of both the field mapping profile and the shower longitudinal profile. It is not clear whether ground coordinates or shower plane coordinates form a better representation for this correlation as on one hand the physics processes are better described in the shower plane, but the method relies on positions and geometries in the ground reference frame.

\begin{figure}[t!]
	\centering
	\includegraphics[width=\linewidth]{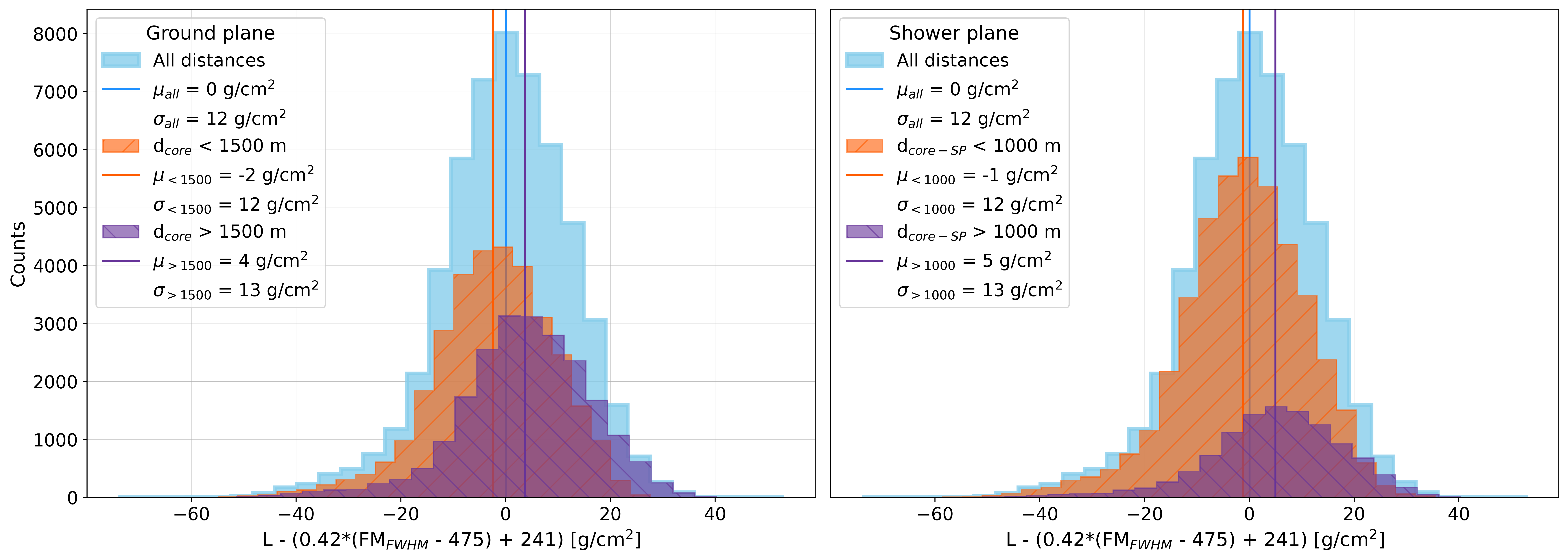}
	\caption{Resolution of the $L$ parameter obtained from the linear fit in Eq. \ref{eq:width_fit} using \FMwidth ~from single observers. In blue we show the resulting residual for all observers. On the left (right) panel, we show in orange the distribution for observers at distances below 1500\,m (1000\,m), and in purple for observers at distances larger than 1500\,m. (1000\,m) in the ground (shower) plane.}
	\label{fig:FWHMreso}
	 \vspace*{.5 cm}
	\includegraphics[width=\linewidth]{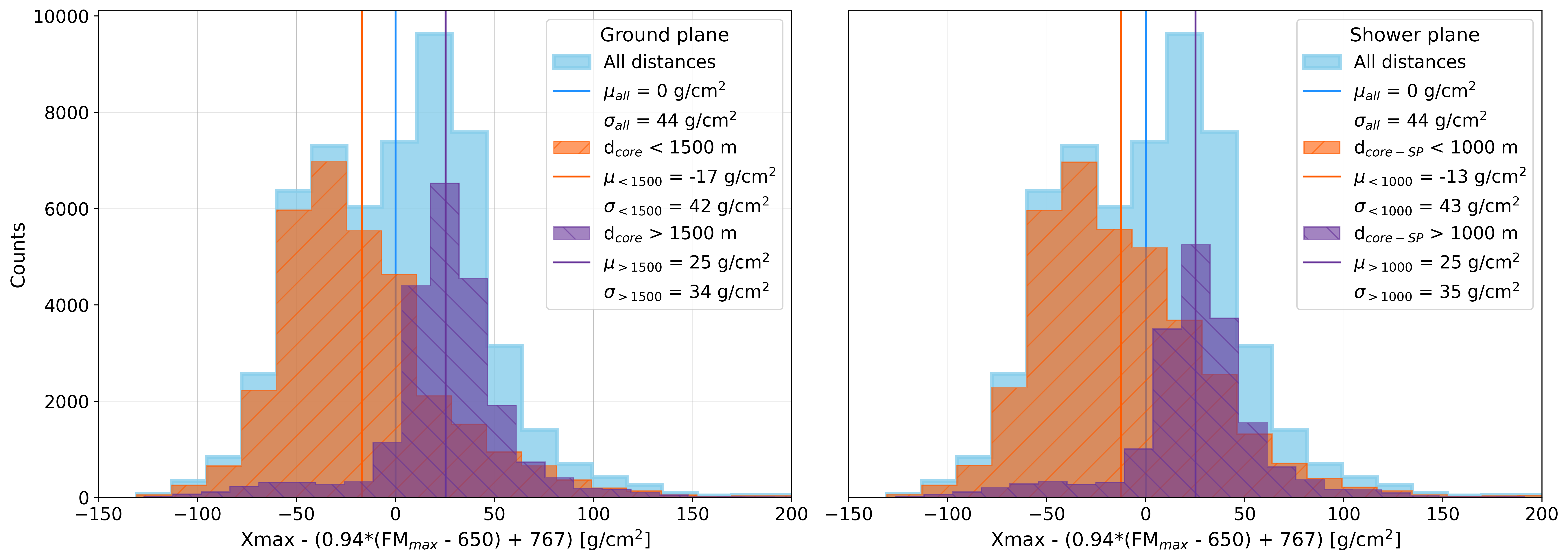}
	\caption{Resolution of the \Xmax ~parameter obtained from the linear fit in Eq. \ref{eq:Xmax_fit} using \FMmax ~from single observers. In blue we show the resulting residual for all observers. On the left (right) panel, we show in orange the distribution for observers at distances below 1500\,m (1000\,m), and in purple for observers at distances larger than 1500\,m. (1000\,m) in the ground (shower) plane.}
	\label{fig:Xreso}
\end{figure}

When using a single observer to estimate $L$ and \Xmax, the radial dependence of \FMwidth ~and \FMmax ~could add a distance dependent bias to the result. Therefore, we further assess the radial dependence by comparing the resolution obtained when considering all observers or only observers close to or far from the core. In Figure \ref{fig:FWHMreso}, we show the distributions of the difference of $L$ from the particle distribution and the one obtained from the fit of the field mapping profile in Eq. \ref{eq:width_fit}, derived using all observers outside the \Cerenkov region. We obtain a resolution of 12\,g/cm$^2$, without significant improvement when separating by the distance of the observer to the core in either reference frame.

In Figure \ref{fig:Xreso}, we show the distributions of the difference of the \Xmax ~from the particle distribution and the one obtained from the fit of the field mapping profile in Eq. \ref{eq:Xmax_fit}, derived from all observers outside of the \Cerenkov region. In this case, using the average correlation for the event is not optimal, as it is clear that there are at least two underlying distributions. If we ignore this, a resolution of 44\,g/cm$^2$ is obtained from the result of a single observer. 
When we distinguish between observers close and far away from the core, either in ground coordinates or in the shower plane, the width of the distribution for remote observers reduces to less than 35\,g/cm$^2$. However, this separation leads to a shift in the distribution expectation value. This can be remedied using different offsets in the calibration curves.

\begin{figure}
	\centering
	\includegraphics[width=.9\linewidth]{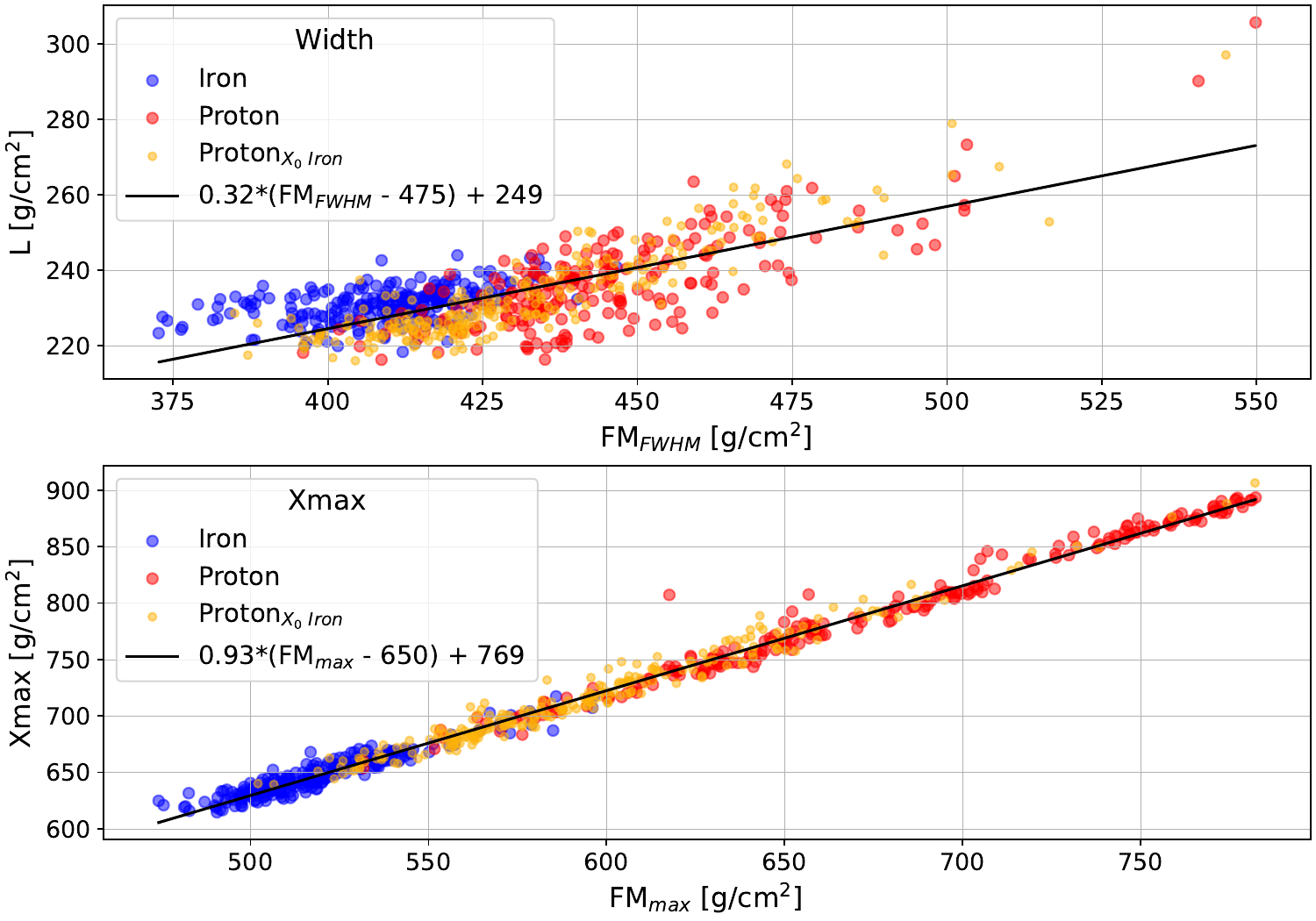}
	\caption{\textbf{Top:} average \FMwidth ~from the field mapping profiles of each proton-induced (red), iron-induced (blue), and proton-induced at same first interaction depth, $X_0$, as iron (yellow), versus the $L$ parameter of the particle longitudinal profile. \textbf{Bottom:} average \FMmax ~from the field mapping profiles of each proton-induced (red), iron-induced (blue), and proton-induced at same $X_0$ as iron (yellow), versus the \Xmax ~of the particle longitudinal profile. The correlation between profiles is quantified by the linear fit (in black) in each panel.}
	\label{fig:primaries}
\end{figure}

The described results were obtained for proton-induced air showers. To probe the primary particle dependence of the field mapping profiles, we performed the same analysis on the iron-induced air showers and on the proton-induced air showers with the same first interaction depth, $X_0$, as the iron-induced showers. In Figure \ref{fig:primaries}, we show the average \FMwidth ~and \FMmax ~over all observers of each simulated event, induced by different primary particle types. Iron-induced field mapping profiles have smaller \FMwidth ~than proton-induced ones, consistent with cascades that develop faster and higher up in the atmosphere and therefore emit the bulk of the radiation in a smaller depth range. A clear trend is seen for \FMmax, where cascades initiated by heavier primaries reach their peak emission higher up in the atmosphere than lighter primaries. For the protons with the same $X_0$ as the irons, \FMmax ~and \Xmax ~are closer to the iron distribution, however still distinguishable.

To assess if the dependence with the particle longitudinal profile of Eqs. \ref{eq:width_fit} and \ref{eq:Xmax_fit} holds for all primary particle types, we performed a linear fit between $L$ and \FMwidth:
\begin{equation}
	L = 0.32~(\text{FM}_{\text{FWHM}}-475\,\text{g/cm}^2) +249\,\text{g/cm}^2,
	\label{eq:width_fit_all}
\end{equation}
with a resulting uncertainty of 0.5 g/cm$^2$ in the $L$ parameter estimate. For \Xmax ~and \FMmax:
\begin{equation}
	\text{X}_{\text{max}} = 0.93~(\text{FM}_{\text{max}}-650\,\text{g/cm}^2) +769\,\text{g/cm}^2,
	\label{eq:Xmax_fit_all}
\end{equation}
with an uncertainty of 0.4 g/cm$^2$ for this \Xmax~estimate. 
When comparing the proton-only fit, in Eqs. \ref{eq:width_fit} and \ref{eq:Xmax_fit}, with the proton and iron fit, in Eqs. \ref{eq:width_fit_all} and \ref{eq:Xmax_fit_all}, we observe that the relation between both the width and the maximum of the particle distribution and the field mapping profile is independent of particle type and interaction depth. However, the values of \FMmax ~and \FMwidth ~are indications for the particle type, similar as the value for \Xmax ~and $L$.

\section{Conclusion}
We have shown that for near horizontal air showers, with a zenith angle of $80\degree$, the time series of the electric field measured at a single observer, located outside the \Cerenkov region, can be used to infer a longitudinal electric field emission pattern, referred to as a field mapping profile. The depth at which the obtained profile reaches its maximum value consistently precedes the particle \Xmax. In addition, we verified that the overall shape of both profiles correlate in both width and depth of maximum.

Assuming that observers have perfect knowledge of the variations of the local electric field, we obtain the average maximum and width from the field mapping profiles of observers outside the \Cerenkov region. These are correlated with the location of particle shower maximum and width, showing that the field mapping profile can be used to infer air shower development. The relation between the field mapping distribution and the particle distribution is universal, therefore the field mapping profile average width and maximum can be used to infer the primary particle type, demonstrating potential for mass composition analysis from radio-frequency measurements.

When using the mapping profile obtained by individual observers, we see a pronounced radial dependence of the correlation between \FMmax ~and \Xmax. Taking this into account, the obtained resolution in the determination of \Xmax ~improves significantly. 

\section*{Acknowledgments}

\noindent
We acknowledge the support from the following funding agencies: Conselho Nacional de Desenvolvimento Cienti\'ifico e Tecnol\'ogico (CNPq); Funda\c{c}ão de Amparo \`a Pesquisa do Estado de Rio de Janeiro (FAPERJ); Coordena\c{c}ão Aperfei\c{c}oamento de Pessoal de N\'ivel Superior (CAPES).

\bibliographystyle{elsarticle-harv} 

\end{document}